\pdfoutput=1
\documentclass[authorversion,nonacm,screen]{acmart}
\usepackage{pdfpages}
\usepackage{subcaption}
\usepackage{cleveref}

\AtBeginDocument{%
  \providecommand\BibTeX{{%
    \normalfont B\kern-0.5em{\scshape i\kern-0.25em b}\kern-0.8em\TeX}}}

\setcopyright{acmcopyright}
\copyrightyear{2018}
\acmYear{2018}
\acmDOI{10.1145/1122445.1122456}

\acmConference[Woodstock '18]{Woodstock '18: ACM Symposium on Neural
  Gaze Detection}{June 03--05, 2018}{Woodstock, NY}
\acmBooktitle{Woodstock '18: ACM Symposium on Neural Gaze Detection,
  June 03--05, 2018, Woodstock, NY}
\acmPrice{15.00}
\acmISBN{978-1-4503-XXXX-X/18/06}

\begin{document}

\title{Multi-Level Quickening: Ten Years Later}
\author{Stefan Brunthaler}
\affiliation{%
  \institution{Nat'l Cyber Defence Research Institute,
    Universit{\"a}t der Bundeswehr, M{\"u}nchen}
  \city{Munich}
  \country{Germany}}
\email{brunthaler@unbiw.de}

\begin{abstract}
  This paper presents important performance improvements for interpreters, exemplified by speedups of up to 5.5$\times$ for CPython.
  Although the original version of this papers was rejected multiple times, the reported speedups have not been achieved by any other interpreter optimization technique since.
  In addition, the paper uses a sound evaluation methodology based on a corollary on Amdahl's law to quantify the speedup potential of benchmarks, which also has not been used in any other paper since.

  This paper documents my best efforts, and includes all of the reviews the paper received, plus some more commentary on my side on what has changed since and what purpose the archived document could serve.
\end{abstract}

\maketitle

\tableofcontents

\paragraph{Outline}
Section~\ref{s:orig-paper} contains the original paper, submitted to the ACM Transactions on Architecture and Code Optimizations, TACO, in 2014.
The prior versions submitted to PLDI'13 (i.e., in November 2012) and PLDI'14 (i.e., in November 2013) were similar in content, but contained fewer material.
These versions are available from the author upon request.

Section~\ref{s:reviews} contains \emph{all} reviews received from PLDI 2013, PLDI 2014, and TACO 2014 in their entirety, with only some markup added to reflect the structure of the reviewing system.

Section~\ref{s:conclusions} contains some comments regarding specific review information.

\section{Preliminary Remarks}

\subsection{History}
The research documented by this paper was carried out in 2011, after relocating from Vienna, Austria, to Irvine, CA, USA in March of 2011.
The gist of the technique, was already mentioned on the last page of my PhD thesis, which postulates the interesting nature of combining two separate techniques of my thesis, namely the use of inline caching through quickening---i.e., the rewriting of interpreter instructions at run-time---with simple static analysis that was implemented to eliminate redundant reference count operations (DLS'12).

As a result of the substantial speedups, the author gave a series of talks about this research in the following universities:
\begin{itemize}
\item October 8th, 2012; Johannes Kepler Universit{\"a}t Linz, Institut f{\"u}r System Software, Prof. Dr. M{\"o}ssenb{\"o}ck.
  \url{http://informatik.jku.at/kolloquium/listTalk.jsp?talkId=284}
\item October 11th, 2012; TU Wien, Institut f{\"u}r Computersprachen, Prof. Dr. Knoop.
  \url{http://www.complang.tuwien.ac.at/talks/Brunthaler2012-10-11}
\item October 11th, 2012; Institute of Science and Technology, Austria, Prof. Dr. Henzinger.
\end{itemize}
These talks coincided with the first submission to the ACM conference on programming languages and implementation (PLDI) 2012.

On April 3rd, 2013, the author was invited to give a talk at Mozilla Research, which was also recorded, broadcast online, and subsequently made available through the Mozilla Research website.
Though the video is not available anymore, the following URL contains the announcement: \url{https://bugzilla.mozilla.org/show_bug.cgi?id=857381}.

\subsection{Other Versions}
At PLDI 2013 in Seattle, in personal conversation Zach Tatlock suggested to formalize the relevant parts and submit to POPL.
After initially doubts, the author began work on formalizing the technique, which led to subsequent submissions to POPL'14, POPL'15, ECOOP'17, and finally to CPP'21, where the paper was finally accepted.

\subsection{Related Work}
In February of 2014, a competing system, ORBIT, was presented at CGO'14.
The author was not aware about this system, but was kindly notified by Christian Wimmer, who attended the conference.
Since there was a surprising amount of overlap between my original paper and the ORBIT paper, the author decided to resubmit quickly to the TACO journal, to establish the independent nature of these discoveries.
This latter step seemed to be particularly pertinent, since a subsequent submission to either PLDI or CGO was impossible with such closely related work.

Many of the ideas presented in this paper resemble parts of the Truffle/Graal ecosystem, but predate the Truffle work by about a year.
An important difference is that multi-level quickening does \emph{not} require a just-in-time compiler, and sidesteps the whole problem of dynamic code-generation altogether.

\newpage
\section{Original Paper}\label{s:orig-paper}
The original paper starts on the following page, no. 4.
The following changes have been made:
\begin{itemize}
\item changed from \texttt{SIGPLAN} to \texttt{acmart} Latex template,
\item removed ACM copyright information;
\item removed ACM subject classification information.
\end{itemize}

\includepdf[pages=-,pagecommand={}]{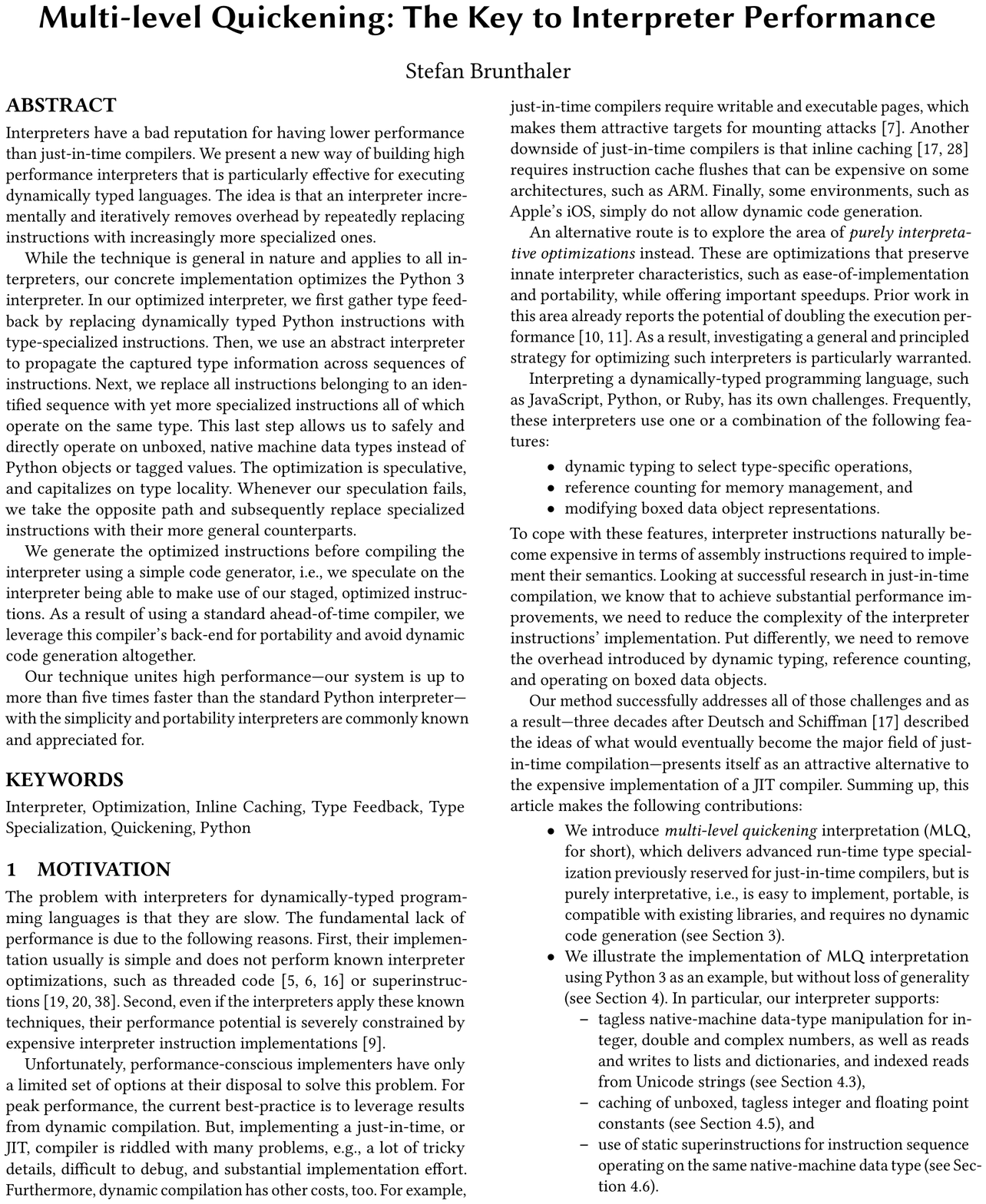}

\section{Reviews}\label{s:reviews}

\subsection{PLDI'13}\label{s:pldi13-reviews}

\subsubsection{Overview}

Before using the abbreviation MLQ, short for multi-level quickening, the project's working title has been ``NAMASTE'', which was the abbreviation for native-machine typed interpreter instructions.
For this reason, these interpreter instructions also carry the \texttt{NAMA\_} prefix.

Reviewers remark on two issues, the compiler error on PowerPC, and the small set of benchmarks.
The compiler error turned out to be a ``fun'' platform quirk of \texttt{gcc}:
My analysis of types used the C type \texttt{char}, since this sufficed to for abstract interpretation of the operand stack.
Unfortunately, however, \texttt{gcc} on Intel machines made \texttt{char}s to be \texttt{unsigned}, but on PowerPC machines, they were \texttt{signed} by default.
After finding this issue, which literally required just to put the \texttt{unsigned} at the right place, everything worked out just fine.
(And taught me the value of using \texttt{uint8\_t} the hard way.)

The second comment regarding the benchmarks was apt, but not really much I could do, as these were pretty much the standard benchmarks for Python performance evaluation at the time.
I remember coming across a larger set of Python benchmarks sometime later on, but could not implement some other benchmarks by myself.
From a reviewer perspective, I did not, and still do not, understand a rejection on such grounds, as it is fairly easy to ask an author to provide more benchmarks.

\subsubsection{First reviewer's review}

\paragraph{Classification}          
B: I can accept this paper, but I will not champion it (accept,
but could reject).

\paragraph{Summary of the submission}        

This paper presents a dynamic compilation technique in which a script
interpreter dynamically replaces instructions between multiple level of
abstraction. Feedback is gathered by replacing Python-Level instructions with
medium level instructions with type information encoded. Type propagation helps
lower these instructions to low level instructions operating on native-machine
data types instead of python objects.  The approach is speculative, so
de-quickening is done when necessary.  The prototype demonstrates real speedup.

\paragraph{Evaluation} 

Points in Favor:

\begin{itemize}
\item While not exceeding the JIT on average, the performance results look
  promising.

\item Complexity is likely to be less than a JIT.

\item   The idea is interesting. The proposed technique quickens a sequence of
  instructions
  into a new sequence (many-to-many optimization) through type propagation.  This
  allows changes of the in-memory representation for some of the values on the
  stack.
  Since the techniques are speculative, de-quickening is done when necessary.
\end{itemize}

Points Against:

\begin{itemize}
\item The paper does not make clear how much less complexity (if any) this approach
  has over a JIT.

\item   The paper argues that language implementers need a viable alternative JITs,
  and that quickening interpreters fit that niche because they require less
  development and maintenance effort, even though they are likely to achieve
  less
  performance. While JITs are expensive to develop and maintain, pervasive and
  free JITs exist. Several projects have identified this opportunity, creating
  Python
  implementations to leverage these JITs: PyPy, JPython, IronPython, Unladen
  Swallow. And although PyPy is a "multi-year, multi-person effort," it's
  incorrect to say that all of that programmer effort went toward targeting
  the Java and .NET VMs. Thus, the claim that JITs require prohibitive
  development costs may be unfair.

\item   The paper proposes non-trivial changes to the interpreter. Specifically, it
  nearly
  triples the number of instruction opcodes, uses two different in-memory
  representations of common data types, and calls for a dynamic recompilation
  system. The total increase in code size is not fully quantified (how big is
  the CPython event loop?), but requires at least 2500 new lines of code (out of
  how many?) and changes to another 2100 lines. I estimate that this complexity
  adds to the development and maintenance costs of the MLQ interpreter. Does it
  contradict the stated goal of "a viable alternative to the costly
  implementation of
  a JIT compiler."
\end{itemize}

Comments for Improvement:

\begin{itemize}
\item Only a small set of small benchmarks are evaluated.  Evaluation more.

\item   http://speed.pypy.org shows performance results for PyPy (trunk) on fannkuch.

\item   The discussion section would benefit from an analysis of the remaining
  performance differences with PyPy. Why does PyPy outperform MLQ most of the
  time?
  What is the remaining inefficiency---is it dispatch cost, or would MLQ need to
  optimize longer sequences?  Support those answers with empirical data.

\item   The paper would benefit from a clear explanation of 'quickening' in Section 2.

\item   Avoid color. Not all printers are color.  The proceedings are not in color.
\end{itemize}
Evaluation:
\begin{itemize}
\item Interesting paper and a novel technique.  At this early stage, more convincing
  needs to be done.
\end{itemize}

\subsubsection{Second reviewer's review}

\paragraph{Classification}

C: This paper should be rejected, though I will not fight strongly
against it (reject, but could accept).

\paragraph{Summary of the submission}
This paper introduces the next step in a applying quickening to give more
efficient interpreters.  Previous work in DLS 10 and ECOOP 10 introduced
quickening, a new instruction format, and the use inline caching to enable
further type-specialized quickening.  This paper introduces the next step,
which builds upon the types observed in the inline cache version to allow for
even more specialization of the instructions and hence more efficient
interpretation.

\paragraph{Evaluation}
I like the general area of investigating how to make interpreters more
efficient. The idea presented in this paper does seem like a nice next step
along the road of quickening.  However, I think that it is important to look at
this paper in terms of the delta from the previous papers on quickening. There
definitely is something new here ... the dynamic profiling and switching to the
even more specialized NAMASTE IR is new.   However, it is a bit hard to extract
from the paper that this is the case.

The experimental results in Figure 9 use the switch-dispatch interpreter as the
baseline and then show the threaded version, the INCA version and the MLQ
versions as speedups relative to the baseline.  Only the MLQ is new, the
previous best was INCA, and the performance of the others is already reported
(on the same benchmarks) in the DLS 10 paper.  Thus, it would seem to me that
the results in this paper should concentrate on the improvements over the
previous best.

I also think, that given that this is the next step in an established line of
research, that the number of benchmarks needs to be increased.  I believe that
this is a relatively small benchmark set (it looks a bit larger because of
using several different input sizes), and each benchmark is also quite small.

Some smaller points,
\begin{itemize}
\item - table 9 should really be normalized on INCA
\item   - in section 3.5 you say you do not need a elaborate undo mechanism because
  you only use NAMSASTE on side-effect free sequences. What is the impact of the
  descision to limit to side-effect free sequences?
\item   - you say you cannot run the INCA interpreter on PowerPC because of
  compilation errors - doesn't that somewhat negate your claim that interpreters
  are great because they are very portable?  Why isn't INCA portable?
\item   - in your comparison on space (end of section 4.2) you first give some
  numbers, 7MB vs 20MB, but then say it is not a fair comparison. If it is not
  something that can be compared, then it shouldn't be compared.
\item   - in related work, the work by Gagnon on SableVM ( CC 2003 ) used instruction
  specialization (what you call quickening) in inline threaded code, so it might
  be worth mentioning. Recent work on specializing on types for JITS was by
  Chevalier-Boisvert (CC 2010).
\end{itemize}

\subsubsection{Third reviewer's review}

\paragraph{Classification}

C: This paper should be rejected, though I will not fight strongly
against it (reject, but could accept).

\paragraph{Summary of the submission}

The paper proposes a new interpreter design that uses three levels (high,
intermediate, low level, ) of instruction sets in executing Python for high
performance. It describes the details of the design of the new interpreter and
made experiments by running six programs of computer languages benchmark game
on both PowerPC and Nehalem and compare performance over cPython Interpreter
and PyPy interpreter. The performance improvements are up to 4X over cPython.
The new interpreter outperforms PyPy for one benchmark program.

\paragraph{Evaluation}

\begin{itemize}
\item + a new innovative design of the interpreter for dynamically typed languages
  and decent performance improvements while maintaining portability.
\end{itemize}
\begin{itemize}
\item - limited set of benchmark programs - more real-world applications with dynamic
  type changes should be tested.
\item   - little information on how to generate low-level instructions except for a
  statement of "We adapted the existing Python code generator of the INCA system
  to generate the C NAMASTE instruc- tion derivatives"
\item   - missing information on how often deoptimizations (undo) occurred for each
  benchmark program
\end{itemize}

\subsubsection{Fourth reviewer's review}

\paragraph{Classification}

C: This paper should be rejected, though I will not fight strongly
against it (reject, but could accept).

\paragraph{Summary of the submission}
The paper describes a bytecode interpreter that performs just-in-time
trace compilation, not to native assembly code like many tracing JITs,
but to an intermediate language named NAMASTE that is roughly at the
same abstraction level as C. During the JIT translation to NAMASTE,
the system performs type specialization, but only for a statically
determined set of type combinations, which includes arrays of floating
point numbers.

\paragraph{Evaluation}
\noindent{}
Pros:
\begin{itemize}
\item The idea of JITing to a portable intermediate language is a good idea.
\end{itemize}
Cons:
\begin{itemize}
\item There are major holes in the description; the paper reads like a
  mystery novel.
\item The idea of only doing type specialization for a fixed set of
  type combinations is not such a good idea, as it misses out on
  many opportunities for optimization.
\end{itemize}

Comments for improvement:

\begin{itemize}
\item * page 1
  \begin{itemize}
  \item     "Next, we use type propagation to further specialize down to
    a low abstraction-level instruction set..."
    This sounds a lot like a JIT, what's the essential difference?

  \item     "for powering much of Internet, and increasingly smartphones,"
    -> missing "the"
    "for powering much of the Internet, and increasingly smartphones,"

  \item     "that is easy to implement, portable, and requires no dynamic code generation"
    But isn't quickening a form of code generation?

  \item     "We automatically generate the instruction sets before compiling the
    interpreter using a standard ahead-of-time compiler,"
    This process is never described in the paper.
  \end{itemize}

\item page 2

  \begin{itemize}
  \item "by rewriting---or quickening"
    Why do you use the term "quickening" when there's already a term for
    this: type specialization?

  \item     "Figure 1 shows the idea: we quicken complete sequences of the original
    instructions to their optimized derivatives..."
    Sounds like a tracing JIT.
  \end{itemize}

\item page 3

  \begin{itemize}
  \item "This is contrary to a JIT compiler, which needs a dedicated back-end
    for each target architecture..."
    Does Namaste have these portability problems too? At this point in the
    paper it's unclear whether and how Namaste differs from assembly code.

  \item     "only inline caching taking place at first"
    The connection between inline caching and type profiling needs to be
    explained!

  \item     "arbitrarily large interpreter instruction set"
    Why does MLQ have an arbitrarily large instruction set?
  \end{itemize}

\item page 4

  \begin{itemize}
  \item "holds the instruction number of target instruction"
    grammar problem

  \item     "The previous section describes how to collect type information..."
    Not really.

  \item     "TARGET(PROF\_JUMP\_ABSOLUTE): ..."
    Is this C code? It would be nice to know what these macros do.
  \end{itemize}

\item page 5

  \begin{itemize}
  \item section 3.3.1
    What about handling the overflow into big integers during an addition?

  \item     "
    else
    PUSH( result.word );
    NEXT\_INSTR();
    "
    Bad indentation.

  \item     "A straightforward way to deal ..."
    This description is too hand wavy.

  \item     "Our implementation follows the first approach and doubles..."
    This approach is wasteful of space and not scalable for languages
    with user-defined unboxed structure types.
  \end{itemize}

\item page 6

  \begin{itemize}
  \item "TARGET(NAMASTE\_FLOAT\_LIST\_SUBSCR)"
    When does this specialized code get created? Is this manual and you
    only optimize certain combinations?
  \end{itemize}

\item page 7

  \begin{itemize}
  \item "Unfortunately, wecouldnot run the INCA interpreter on our PowerPC
    system, because of compilation errors."
    This discredits your portability claim.
  \end{itemize}

\item page 8

  \begin{itemize}
  \item "We adapted the existing Python code generator of the INCA
    system to generate the C NAMASTE instruction derivatives"
    This needs to be explained... should have been explained earlier in the paper.

  \item     "written in C"
    Say this earlier!
  \end{itemize}

\item page 9

  \begin{itemize}
  \item "other candidates for start instructions that we do not currently
    support, such as LOAD\_ATTR, LOAD\_NAME, LOAD\_GLOBAL, LOAD\_DEREF"
    These seem pretty important, especially LOAD\_ATTR.

  \item     "This is in contrast to the Python work mentioned above, which
    generates instruction derivatives statically and thus leverages the
    ahead-of-time compiler when compiling the interpreter"
    What does MLQ do? Does it use the same approach as Brunthaler?
    How does MLQ determine which specializations to generate?

  \item     "the core idea behind multi-level quickening interpretation:
    ... pre-compile several alternative templates for the same code sequence
    and choose between them at run time"
    Say this earlier!
  \end{itemize}
\end{itemize}

\subsubsection{Fifth reviewer's review}

\paragraph{Classification}

B: I can accept this paper, but I will not champion it (accept,
but could reject).

\paragraph{Summary of the submission}

This paper describes Multi-Level Quickening (MLQ), a technique to improve the
performance of an interpreter for dynamically typed languages.
MLQ reduces the overhead of dynamic typing and boxing/unboxing by rewiting (or
quickening) bytecodes dynamically at runtime based on profile information.
The results showed MLQ gave 1.5x to 4x performance improvement over the default
CPython 3.2.3 on Intel Nehalem in small benchmarks.

\paragraph{Evaluation}

Overall, this paper is well written and the motivation of MLQ is very clear. I
enjoyed reading the paper.
MLQ is simple but effective to eliminate runtime overhead in interpreters for
dynamic languages.

I feel that the biggest weakness of this paper is its evaluation.
The author(s) uses only small benchmarks for the evaluation. Though I
understand that this is because there is no major large benchmark for Python 3,
I really want to see the results for realistic workload to confirm that the MLQ
really improves the performance of real-world programs.

I feel that the related work section could be more complete by discussing the
following paper, which presents a technique to reduce boxing/unboxing overhead
in pypy's tracing JIT. They aim the same goal of reducing the boxing overhead
by specializing the execution path though they used the dynamic compiler
instead of interpreter for execution.
\begin{itemize}
\item - Carl Friedrich Bolz, Antonio Cuni, Maciej FijaBkowski, Michael Leuschel,
Samuele Pedroni, and Armin Rigo. 2011. Allocation removal by partial evaluation
in a tracing JIT. In Proceedings of the 20th ACM SIGPLAN workshop on Partial
evaluation and program manipulation (PEPM '11).
The object allocation removal techniques based on escape analysis in
(method-based) JIT compilers are also interesting to discuss as related work.
\item An example of the papers on escape analysis is
- Ajeet Shankar, Matthew Arnold, and Rastislav Bodik. 2008. Jolt: lightweight
dynamic analysis and removal of object churn. In Proceedings of the 23rd ACM
SIGPLAN conference on Object-oriented programming systems languages and
applications (OOPSLA '08).

\end{itemize}

\textbf{Question:}
The paper claims that this technique can be applicable to other languages.
However, it might be much difficult to rewrite bytecodes at runtime safely if
the language supports multi threading (e.g. Ruby). Do you have a good idea on
how to implement the quickening efficiently in multi-threaded environments?

\newpage
\subsection{PLDI'14}\label{s:pldi14-reviews}

\subsubsection{First reviewer's review}

\paragraph{Classification}
C

\paragraph{Summary of the submission}

The authors present an optimized interpreter for Python. It essentially
compiles high-level bytecode into lower-level typed bytecode speculatively and
falling back on the original when needed. The paper demonstrates improvements
over the baseline CPython interpreter ranging from very small to ~4x in a
couple cases.

\paragraph{Strengths}
The paper tackles one of the most heavily used interpreters - the standard
Python interpreter. It demonstrates nice speedups there (up-to 4x) on certain
codes. The technique should not interfere with the usage of popular Python
libraries (which often use native code underneath).

\paragraph{Weaknesses}

CPython is a fairly poor baseline for performance. There is no comparison with
other Python systems that are arguably similar - Jython, IronPython - that they
are implemented over a lower-level typed (but other substantially different)
bytecode.

\paragraph{Evaluation}

This paper investigates improving the performance of the standard Python
interpreter through a technique called quickening. The authors created an
optimized interpreter that provides type-specialized variants to standard
Python bytecodes. Python bytecode is analyzed to propagate type information and
replace generic bytecodes with type specific ones. Common sequences of
type-specialized bytecodes are lowered further to remove redundant checks. The
lowering can be speculative - the system will fall back on the original generic
bytecodes if speculation fails.

The lower-level instructions are created in a "staging" step. The paper doesn't
really go into enough detail on how this is done. It appears this is ahead of
time, and a new optimized interpreter is created. I'm not sure how it is
trained / bootstrapped or what the impact is on interpreter size.

The second step appears to be a bytecode precompilation step - though perhaps
some of this could be done just-in-time on execution. This step lowers the
general bytecode to specialized bytecode. It uses speculative type propagation
(via abstract interpretation) to drive this. It is able to specialize for types
including integers, floats, strings, and lists.

The evaluation demonstrates speedups on Python benchmarks - mostly modest, but
in some cases a factor of 2-4x over the baseline interpreter.

I found the paper somewhat confusing to read. I'm not clear on when/where the
staging and quickening steps happen. A more detailed architecture section would
be very helpful in this regard.

It would also be helpful to see comparisons with Jython and IronPython - I
expect they run all the benchmarks listed here.

Is there any impact on startup time or memory footprint with this system? Both
are often important considerations for interpreters.

Finally, the system has a fair bit of complexity - it's not really clear this
is a worthwhile step instead of moving to a simple JIT.


\subsubsection{Second reviewer's review}

\paragraph{Classification}

B

\paragraph{Summary of the submission}

The paper presents a strategy to improve performance of interpreters based on
profiling, type inference, and the use of type-specialized byte code
instructions. The strategy was implemented as an extension to the Python 3.3.2
interpreter and evaluated using a few codes with good results.

\paragraph{Strengths}

The proposed strategy of improving the interpreter for performance makes sense
since extending the interpreter is not only easier than adding a JIT compiler,
but it also enables portability to new instruction sets.

\paragraph{Weaknesses}
The main drawback of the paper is that the strategy of using unboxing, guards,
and type-specialization is not that original. Also, the evaluation section
needs improvement in the presentation.

\paragraph{Evaluation}

The proposed strategy of type specialization is not tremendously novel. This
does not mean that the paper is not interesting since it has experimental
value. It shows how well this strategy applies to the Python interpreter.
The overall presentation of the paper is good, but it seems that some
clarifications are needed. Here are a few suggestions:
\begin{enumerate}
\item 1.      The reference to inline caching in the introduction to section 4. is not
  clear. It needs at least a reference and better a short description of what it
  means.
\item   2.      On page 3, the INCA\_LONG\_ADD function has parameter type
  (PyLongObject.PyLongObject.S). A short description of what this type means
  would be useful.
\item   3.      The whole reason behind Section 4.3.2 is not clear. Why do casting? I
  believe that casting does not arise naturally in Python. Is this assumption
  wrong?
\item   4.      Figure 3 needs an explanation. Right now the figure is only mentioned in the
  text.
\item   5.      The reference to Andahl's law in Section 5.2 is not clear. In the end I
  don't know how maximum speedup is computed. Is the computation of numerical
  values assumed to take zero time?
\item   6.      Are superinstructions used? What is the effect of the superinstructinos?
\item   7.      What does it mean to have only INCA? Is it just avoiding the type check at
  the beginning of the generic arithmetic instruction?
\item   8.      Would it be possible to correlate maximum speedup to frequency of arithmetic
  operations in the program of Table 3?
\end{enumerate}

\subsubsection{Third reviewer's review}

\paragraph{Classification}

B

\paragraph{Summary of the submission}

This paper presents a multi-level quickening optimization for interpreters.
General interpreter instructions are replaced by type specific ones, and they
can be reverted when the speculation fails.

\paragraph{Strengths}

The approach improves performance while maintaining the interpreter's core
properties.
The speedup is up to 450

\paragraph{Weaknesses}

The evaluation is insufficient.
Many of the benchmarks cannot gain enough performance improvement, which limits
the applicability?

\paragraph{Evaluation}

There should be more detailed evaluation on the individual techniques and
parameters of the approach.
For example:
\begin{enumerate}
\item 1. Doubling the operand stack size enables the optimization for complex
  numbers, but does it incur more cache misses? (possibly not, but prove it)
\item   2. How often does the type speculation fails, and how much is the cost of the
  deoptimization?
\item   3. How did you select the thresholds in Figure 3? I think they would affect the
  misspeculation rate.
\item   4. What is the performance contribution from techniques in section 4.5 and
  4.6?
\end{enumerate}
Table 3 lists 29 benchmarks, but only 10 of them are evaluated (Figure 4),
including ones with minor improvements.
The other 19 benchmarks are not evaluated because they have little potential,
but do they have any slowdown?

In section 5.2, you collected the interpreter's execution time by perf tools.
Did you assume an ideal speedup for the interpreter, so that you could
calculate the maximum possible speedup (potential) with Amdahl's law?
If so, is the ideal speedup a pre-defined fixed value or infinity?

In section 5.4, you mentioned the overall *efficiency*.
From the context, I guess it is the actual speedup divided by the potential
one, but this should be clarified in the paper.

In Figure 1, should the last solid line be a dotted one, from STORE\_FAST to
NAMA\_FLOAT\_MULT?


\subsubsection{Fourth reviewer's review}

\paragraph{Classification}

B

\paragraph{Summary of the submission}
Interpreters typically do not used typed based optimizations. The paper
presents a multi-level optimization scheme for interpreters. In the first
level, instructions that express typed operations are presented. Optimization
to the second level allows unboxed operands (hardware primitive types) to be
consumed and produced by operations within a basic block, further reducing the
overhead.

\paragraph{Strengths}

The paper describes an interesting and important problem and provides a
complete system that solves the problem. Significant speedups are achieved on
the targeted Python programs.

\paragraph{Weaknesses}

The paper lays out the trade-offs between interpreters and dynamic compilers:
interpreters are more secure, portable and easier to develop. Compilers are
faster. Part of the goal of the interpreter was to allow faster performance to
to reduce the advantage of dynamic compilers in this trade-off. Having a
quantitative comparison vs. a dynamic compiler would be useful.

The paper is largely (but not entirely) a well-engineered collection of
techniques that existed in (often) less powerful forms in other systems.

\paragraph{Evaluation}

The paper is overall well written and very clear. Concepts are explained well
and the necessary background is provided.

The paper begins with a description of why interpreters are good and why
dynamic compilers (hereafter called compilers) are not good. The main drawback
of interpreters is that they offer slower performance than compilers for most
programs. The paper then describes a two level optimization strategy that gives
good speedups (close to 4 in programs that have high interpretation overheads)
and that mitigates this advantage of compilers. It would be good to know how
close the MLQ systems performance is to, e.g., Numba or PyPy's performance.
Clearly, the closer MLQ is the greater the reason to use interpreters. In
fairness, almost all dynamic compilation systems start with an interpreter
phase so speeding that up (and gathering the information that MQL will gather)
improves even compiler based systems, as the paper notes.

I have a second question related to performance. The IBM Nnija project
(Moreira, et al.) designed special Complex classes that could be easily unboxed
by a compiler in strings of operations without global side effects. The
intention of this was to allow arithmetic to be performed on objects without
creating objects to hold temporaries. This made a significant performance
improvement. I would think the same is true of your system. Do you have a feel
for how much of your performance improvement comes from knowing types, how much
from not unboxing, how much from not having spurious object created on the
heap, how much from not having to have the operation check for the specific
type each time, etc.

Are there issues with multithreading? It appears not since when you unbox you
make a local shared copy and since this is necessarily in straight-line code
you necessarily keep something that might be shared across threads in, e.g.,
CPtython, private for a relatively short period of time.

In the Amdahl's law discussion, you talk about the "top-quadrant" of Table 3,
but Table 3 is divided into either 3 or 6 parts, not 4.

\newpage
\subsection{TACO'14}\label{s:taco14-reviews}


As mentioned before, I renamed the paper from NAMASTE to MLQ to focus on the essence of the technique.
Besides fixing the PowerPC issue (also mentioned above), I did manage to put a lot more thought and work into carefully designing a proper evaluation methodology.
In addition, I put in some more effort in optimizing native-machine mapping of compound, non-scalar data types, such as lists and dictionaries.

\subsubsection{Editor's comments}
As the referees describe, a more complete submission (particularly in
terms of experiments) is needed to better evaluate the ideas in the
submission.  It is possible that a new submission that reports the
results of these experiments and addresses the referees' concerns will
be accepted by TACO.

\subsubsection{Referee 1}

\emph{Recommendation}: Needs Major Revision

\paragraph{Comments}:
\begin{enumerate}
\item 1. Related work. It seems that ORBIT has no native code generation, either. The paper of ORBIT mentioned the native code generation as a future work.

\item   2. Related work. It’s not required to use a paragraph to show MLQ predates ORBIT and Truffle. The citation of your previous work is clear. The section is used to introduce the related work in this domain, and address the differences.
\end{enumerate}

\emph{Additional Questions}:
\emph{Review's recommendation for paper type}: Full length technical paper

\emph{Should this paper be considered for a best paper award?}: No

\emph{Does this paper present innovative ideas or material?}: Yes

\emph{In what ways does this paper advance the field?}: Explored the new space that the optimization of dynamic scripting language focuses on pure interpreter level optimization.

\emph{Is the information in the paper sound, factual, and accurate?}: Yes

\emph{If not, please explain why.}:

\emph{Rate the paper on its contribution to the body of knowledge in architecture and code optimization (none=1, very important=5)}: 3

\emph{What are the major contributions of the paper?}:
\begin{enumerate}
\item 1. Introduced an on-the-fly byte-code rewriting approach (called MLQ, Multi-Level Quickening by the author) to do specialization in the interpreter of a dynamic language.

\item   2. Gave the reference MLQ implementation for the interpreter of Python 3
\end{enumerate}

\emph{Rate how well the ideas are presented (very difficult to understand=1 very easy to understand =5)}: 3

\emph{Rate the overall quality of the writing (very poor=1, excellent=5)}: 3

\emph{Does this paper cite and use appropriate references?}: No

\emph{If not, what important references are missing?}:
Other Python optimization work, for example "On the benefits and pitfalls of extending a statically typed language JIT compiler for dynamic scripting languages", OOPSLA 2012.

\emph{Should anything be deleted from or condensed in the paper?}: Yes

\emph{If so, please explain.}: Sec 4.4, may use more generic pseudo code.

\emph{Is the treatment of the subject complete?}: No

\emph{If not, What important details / ideas/ analyses are missing?}:
\begin{enumerate}
\item 1.  For completeness, more background about the optimization of dynamic scripting languages should be included. There are a lot of research work in optimizing SELF, JavaScript, Python, Lua, etc.. A high level summary may be required to distinguish this work’s progress.

\item   2. Many details in Sections 4 are dependent on the CPython’s implementation. It’s better to include some related CPython design, such as object format. For example, in order to understand sec 4.4, the reader should know the CPython’s local variable look-up mechanism.
\end{enumerate}

\emph{What type of paper is this?}: This is an extended version of a previously published conference/symposium paper with significantly (generally > 30\%) new material

\paragraph{Recommendations to the authors}
\emph{Please list some concrete actionable items to improve the paper. When the paper is resubmitted, the authors will at least have to explain how they dealt with these recommendations.}:

\begin{enumerate}
\item 1. You may define the term JIT in the context, since JIT is not only limited to generate machine native code. JIT only means on-the-fly compilation. It also could refer to translation to another format, for example byte-code to byte-code.

\item   2. It's hard to understand  "gather type feedback by replacing dynamically typed Python instructions with type-specialized instructions" without reading the author's previous work on Inline Caching with quickening. The replacing action itself cannot provide the type information. It's better to explain the detail in this paper.

\item   3. It is not clear when/under which condition the optimization of the first/second level quickening is triggered.

\item   4. The first level transformation is coupled with inline cache. The paper may explain an alternative way to do the rewriting if the language has no inline cache, or it may mention the limitation of MLQ.

\item   5. MLQ is only inside a basic block. So after the basic block boundary, will all the unboxed objects be boxed? It’s better to explain it, and also show the pros and cons.

\item   6. Section 4.7, Specialization fails. Suppose there are two basic blocks A -->B. If the specialization fails in A, the interpreter triggers the de-optimization of A. Will the de-optimization of B also be triggered?

\item   7. Experiment. Does MLQ encounter specialization failures in the performance evaluation? If yes, please include the analysis of the specialization failure rate. If no, please explain why.
\end{enumerate}

\emph{Please help ACM create a more efficient time-to-publication process: Using your best judgment, what amount of copy editing do you think this paper needs?}: Moderate

\emph{Most ACM journal papers are researcher-oriented. Is this paper of potential interest to developers and engineers?}: Maybe

\subsubsection{Referee 2}
\emph{Recommendation}: Needs Major Revision

\emph{Comments}:
Although the paper has many positive points, there are some additional shortcomings. These are:
\begin{enumerate}
\item (1) The work claims or implies that it is applicable to a wider range of interpreters than the data supports.
\item (2) The work claims a high level of novelty that is neither necessary for publication, nor supported by the prior work in the area.
\item (3) It is not clear why the instruction specialization starts with arithmetic instructions and then propagates the type backwards to earlier executed loads. In other similar systems, such as that of Williams  (CGO 2010), specialized versions of the load instructions are inserted when the loads are first executed.
\end{enumerate}

With respect to scope, the title of the paper claims that multi-level quickening (MLQ) is the key to interpreter performance. Even if we leave aside the question of whether the reported speedups are truly the effect of MLQ rather than simply lucky code placement, there is no evidence to suggest that MLQ is the key to interpreter performance. In truth, interpreters are strongly subject to the Anna Karenina principle: you can make an interpreter slow by getting any one of several things wrong; to make a fast interpreter you need to get all these things right. There is no one key to interpreter performance.

A second, more specific problem with this claim is that MLQ is primarily applicable to dynamically typed languages. If you are building an interpreter for a statically-typed language, MLQ will not help at all. This simple truth stands in stark contrast to the claim in the abstract that "the technique is general in nature and applies to all interpreters". The abstract and introduction have several examples of such hyperbole that a knowledgable  reader will recognize as exaggeration, but will mislead those who have less background in the area.

A similar problem arises when we consider the type of VM bytecode that might benefit from the technique. The examples in the paper all involve propagating types from loads to the stack to computational operations to stores to the stack. In fact, as shown in Table 1, a type propagation sequence ends at a store to the stack. In other words, the approach appears to primarily optimize sequences consisting of LOAD(S), OPERATION, STORE. In a VM that uses virtual registers rather than a virtual stack, this sequence would be a single VM instruction, and there would be no need to propagate types between instructions because within a single instruction it would not be necessary to check the type of any operand more than once. So it appears that single-level quickening is enough for register VMs, and MLQ is useful only where simple operations are broken into a number of VM instructions, as in stack VMs.

Another similar question arises around boxed implementations of dynamic types as compared with tagged implementations. A number of the techniques in the paper are aimed specifically at caching unboxed values. Are the techniques in the paper likely to be just as effective for VM interpreters that use tagged types instead? It's difficult to say, and the paper doesn't provide any evidence either way.

So in fact the statement that MLQ is applicable to all interpreter types is deeply misleading because it is only likely to be beneficial for very specific types of interpreter. It is also highly misleading to claim that MLQ is the key to interpreter performance. The purpose of these points is not that the work in the paper is bad. It's actually very interesting. But the claims made for the work are exaggerated and misleading.

Another problem with the paper is that it claims a degree of novelty that is misleading and arguably untrue. Part of the problem is that the claims are made vague by qualifying words that have no clear meaning. For example "MLQ is the first GENERAL technique that shows how to perform purely interpretive ADVANCED type specialization". The words "general" and "advanced" are too vague for anyone to judge whether or not this statement is true. The paper should contain statements that are clearly true as supported by evidence, rather than statements that are simply difficult for the reader to prove false.

Let us instead consider whether it is true that MLQ is the first technique to perform purely interpretive type specialization. My first response is that it's not terribly important whether previous interpretive type specialization are purely interpretive or part of mixed interpreter/JIT system. To my mind the important question is whether the techniques can be used in a purely interpretive environment. Secondly, it is simply false that no earlier purely interpretive system did type specialization. For example, Williams et al. (CGO 2010) described a fully interpretive system that build a linked representation of the code, where each node represents a specialized version of the original instruction. The interpreter traverses this structure and interprets the opcode in each node. (The current paper incorrectly describes this work as using a separate compiler at run time, however that was a different paper by Williams which appeared in LCPC 2009). Furthermore modifying a VM instruction multiple times during execution is by no means novel. For example Williams (CF 2009) uses a scheme that modifies branch VM instructions for taken/not taken directions. No doubt there are many similar schemes used by others, probably including the author of the current paper. The idea of specializing VM instructions using run-time information is probably almost as old as VM interpreters themselves. The important thing about the current paper is not that it's the first paper to do purely interpretive instruction specialization (it is not the first such paper); the important thing is the particular way in which the current paper does a well-known transformation.

The third additional shortcoming of the paper is that it is entirely unclear why it should be necessary to propagate types from arithmetic instructions back to their corresponding loads in a stack machine. In other type specialization systems, the loads would be specialized when they are executed. Once the load has been specialized it may make sense to propagate the types forwards to the arithmetic instructions. But failing to specialize the loads when they are executed and then building an elaborate system that propagates the types back to those same loads seems illogical. This must be explained.

\emph{Detailed comments}:
\begin{itemize}
\item Abstract and introduction: It would be helpful to turn down the marketing, tighten the claims, and avoid making claims that are only true in a legalistic sense.

\item   Section 2:
  If you show the operand bytes in the example, it will be easier for the reader to recognize this as stack code.

  At the end of section 2 you state clearly that the Python VM operates exclusively on boxed types, even for simple types such as integers. It would be helpful to have this clear statement before the discussion of reference counting. Otherwise readers familiar with type-tagged representations will not understand why there might be a reference count for a simple integer.

\item   Section 3:
  Presuming that the Python opcode is a single byte, is the limit of 256 opcodes a problem? How many instructions are there in the Python VM, and how many specialized opcodes do you create?

\item   Section 4:
  Fig. 3 is too small to read. The dotted font is particularly hard to read.

\item   Section4.2:
  Tell us what the .-operator does before showing the example.

  The description of "any sequence of instructions that has no side-effects" is unclear. Surely the purpose of any instruction is to change the state of the VM. So how can any instruction have no side-effects?

\item   Sections 4.3.1 and 4.3.2
  These could be shortened significantly. Anyone familiar with the implementation of stack VMs will be familiar with the using unions to allow a variety of types on the stack.

\item   Section 4.3.3
  It is worth reminding the reader here that using the regular Python representation of a complex number will require only one stack slot, because the stack slot will contain a pointer to the structure containing the complex number. The problem arises when we unbox the complex number, and have to put two items on the stack.

  Change "more standard types" to "additional standard types".

  At this point it might also be worth mentioning stack caching. Stack caching boxed items is probably much simpler than stack caching unboxed items.

\item   Section 4.4
  Keeping the list size in a local variable. You should make clear that this is a variable that is local to the C interpreter function, not local to the Python function. A variable that is local to the interpreter is, effectively, global to the Python program. How is it possible to make sure this variable has the correct value at all times? For example, if the code consists of two NAMA\_LOAD\_LIST instructions back-to-back. We pop the topmost stack item, and now the cur\_list\_size local variable has the wrong value. How does the interpreter know that the value is wrong?

\item   Section 4.5
  You should briefly discuss whether these optimizations are useful for a register VM.

\item   Section 5.1
  "We [verb missing] the frequency"

\item   Section 5.2
  The geometric mean is an unusual choice when dealing with multiple runs of the same benchmark. Outliers are almost always highly asymmetric when timing programs (you get large outliers on the high side of execution times, but hardly any outliers on the low side).

  "50\% speedup". The term speedup is often misused, and although you use it correctly here I think there is a danger of confusion. On the other hand everyone knows what a 1.5x speedup is.

\item   Section 5.4
  PyPy3 seems to give much worse performance than regular PyPy. Why do you not compare with PyPy rather than a beta version of PyPy3 that seems to have huge performance problems?

\item   Section 6
  It's not clear to me that the historical reference note adds much. No doubt the researchers who developed ORBIT and Truffle also wrote down their ideas and gave talks along the way.
\end{itemize}

\emph{Additional Questions}:
\emph{Review's recommendation for paper type}: Full length technical paper

\emph{Should this paper be considered for a best paper award?}: No

\emph{Does this paper present innovative ideas or material?}: Yes

\emph{In what ways does this paper advance the field?}: The paper presents another variation of VM instruction specialization in interpreters for dynamically typed languages. The idea of dynamic instruction specialization in interpreters is not novel. But the particular techniques proposed are interesting, as is the implementation and experimental evaluation.

\emph{Is the information in the paper sound, factual, and accurate?}: No

\emph{If not, please explain why.}: To be more precise, much of the information is factual and mostly accurate but some of the claims are exaggerated or misleading, and some fundamental data is missing.

\emph{Rate the paper on its contribution to the body of knowledge in architecture and code optimization (none=1, very important=5)}: 3

\emph{What are the major contributions of the paper?}:
\begin{itemize}
\item - A new variation on instruction specialization for interpreters for dynamically-typed languages
\item   - A good implementation of the proposed technique in an interpreter for Python
\item   - An experimental evaluation that is interesting, even if it neglects to present what is perhaps the key to understanding the effectiveness of the technique
\end{itemize}

\emph{Rate how well the ideas are presented (very difficult to understand=1 very easy to understand =5)}: 4

\emph{Rate the overall quality of the writing (very poor=1, excellent=5)}: 4

\emph{Does this paper cite and use appropriate references?}: Yes

\emph{If not, what important references are missing?}: The paper cites appropriate references but the description of at least one piece of existing work papers is difficult to reconcile with the contents of the referenced paper.

\emph{Should anything be deleted from or condensed in the paper?}: Yes

\emph{If so, please explain.}: Sections 4.3.1 and 4.3.2 occupy more than a page but deal with very standard techniques found in almost any interpreter. These subsections could be shortened. It would also be helpful to the reader to know earlier in the paper about the boxed representation of all variables in the Python interpreter, as compared to type-tagged variables in many other implementations of dynamic languages.

\emph{Is the treatment of the subject complete?}: No

\emph{If not, What important details / ideas/ analyses are missing?}: It is well known that indirect branch prediction plays a major role in the performance of many interpreters. The results presented in section 5.3 strongly suggest that what the author is measuring is a second-order effect on indirect branch prediction rather than the primary effect of instruction specialization. But the author does not appear to have investigated this possibility, although it is the explanation that is best supported by previous studies on interpreters.

\emph{What type of paper is this?}: This paper is an original, previously unpublished, paper (to the best of my knowledge)

\paragraph{Recommendations to the authors}
\emph{Please list some concrete actionable items to improve the paper. When the paper is resubmitted, the authors will at least have to explain how they dealt with these recommendations.}: This is an interesting paper that I enjoyed reading and learned something from. Much of the paper is good and I hope to eventually see it in print. However, there are some shortcomings that need to be fixed before this can happen.

It is well known that there are three main costs in bytecode interpretation:
\begin{itemize}
\item (1) Dispatch (fetch opcode and jump to handler code)
\item   (2) Operand access
\item   (3) Performing the actual computation
\end{itemize}
Pretty much all prior research shows that for statically-typed languages with fine-grained VM instruction sets, that interpreter instruction dispatch is the largest of these costs. The main reason is the poor performance of real-processor indirect branch predictors on the interpreter's dispatch branch.

For dynamically typed languages there are other important interpreter overheads, but there is no clear evidence that indirect branch prediction is unimportant for these types of interpreters. (For a contrary view see Rohou et al. 2013)

There has been prior work on interpreters that attempted to reduce i-cache misses (Brunthaler 2011) in interpreters for dynamically typed languages by changing the layout of the code that implements the VM instructions. However, McCandless (2011) showed conclusively that the technique had no effect on i-cache misses, and the performance effect was entirely due to arbitrary effects on indirect branch prediction. Given that indirect branch prediction appears to be the most common reason for the effectiveness of interpreter optimizations, we should always consider the possibility that the apparent effectiveness of any new technique is simply the result of a lucky interaction with the indirect branch predictor.

It is well known that interpreter VM instruction specialization can have an arbitrary impact on indirect branch prediction (Casey et al. 2005). There are at least three effects: (1) more versions of the code that implements a VM instruction leads to more separate dispatch branches in (token) threaded interpreters; (2) multiple versions of the code that implements a VM instruction located at different points in memory can give more information to a two-level indirect branch predictor (McCandless 2011); (3) Multiple version of the code that implements an instruction can result in more possible targets for the dispatch branch, which can result in worse branch prediction.

Given this well-known result, one would expect that a paper on instruction specialization for interpreters should consider whether the positive results are actually from the proposed techniques, rather than simply from a lucky interaction with the indirect branch predictor.

Section 5.3 deals briefly with hardware performance counters. It reports that the reduction in (executed? retired?) branches is proportional to the speedup. However, there is a 29-fold reduction in branch mispredictions, and a 12-fold reduction in stalled back-end cycles.

These results suggest that the speedup from the author's techniques is partly the result of a reduction in executed instructions and branches. However, it also appears that a large part of the speed improvement comes from the remaining branches becoming much more predictable.

An obvious question is whether the reduction in branch mispredictions is the result of type checking branches being removed or the result of the dispatch branch being better predicted.

Clearly part of the reduction in branch mispredictions might be explained by type checking branches being removed. However, there is no reason to think that removing one type checking branch might render the others more predictable. On the contrary I would expect that type checking branches would be strongly correlated, allowing a two-level predictor to improve its prediction of the next type checking branch. I would therefore expect that removing type checking branches would leave the remaining branches less predictable rather than more. So this explanation does not seem consistent with the data.

The other likely explanation is that the reduction in branch mispredictions is due to the interpreter dispatch branch(es) becoming more predictable, because of the existing well-known effects of interpreter specialization on indirect branch prediction. If this is the reason for the reductions in branch mispredictions, it is likely to be an effect that greatly depends on the luck of how the C compiler lays out the interpreter code. This explanation seems consistent with the measured data.

The above two explanations are not the only possible ones. But given that the most likely explanation for the branch prediction improvement appears to rely on an arbitrary artifact of the layout of the interpreter code by the C compiler, it is something that must be investigated and explained thoroughly before the paper can be considered ready for publication. An obvious first step would be to look at the indirect branch native performance counters.

\emph{References:}

\begin{itemize}
\item Erven Rohou et al. Branch Prediction and the Performance of Interpreters – Don’t Trust Folklore. INRIA Research Report 8405, 2013.

\item   Stefan Brunthaler: Interpreter Instruction Scheduling. CC 2011: 164-178

\item   Jason McCandless, David Gregg: Optimizing interpreters by tuning opcode orderings on virtual machines for modern architectures: or: how I learned to stop worrying and love hill climbing. PPPJ 2011: 161-170
\end{itemize}

\emph{Please help ACM create a more efficient time-to-publication process: Using your best judgment, what amount of copy editing do you think this paper needs?}: Moderate

\emph{Most ACM journal papers are researcher-oriented. Is this paper of potential interest to developers and engineers?}: Maybe

\subsubsection{Referee 3 -- David Ungar}

\phantom{bingo}

\emph{Recommendation}: Reject

\emph{Comments}:
see attached file [Stefan Brunthaler: reprinted here with David's permission!]

\emph{Additional Questions}:
\emph{Review's recommendation for paper type}: Full length technical paper

\emph{Should this paper be considered for a best paper award?}: No

\emph{Does this paper present innovative ideas or material?}: Yes

\emph{In what ways does this paper advance the field?}: It presents a new methodology for optimizing an interpreter, as far as I can tell.

\emph{Is the information in the paper sound, factual, and accurate?}: No

\emph{If not, please explain why.}: Unsound evaluation methodology, failure to consider both key sources of overhead and alternative architectures such as meta-circular JITTers. See full comments below.

See attached file.

\emph{Rate the paper on its contribution to the body of knowledge in architecture and code optimization (none=1, very important=5)}: 4

\emph{What are the major contributions of the paper?}: A method for rewriting bytecodes

\emph{Rate how well the ideas are presented (very difficult to understand=1 very easy to understand =5)}: 3

\emph{Rate the overall quality of the writing (very poor=1, excellent=5)}: 2

\emph{Does this paper cite and use appropriate references?}: No

\emph{If not, what important references are missing?}: Jikes/RVM work
See attached file.

\emph{Should anything be deleted from or condensed in the paper?}: Yes

\emph{If so, please explain.}: Excessive claims for performance vs complexity: see full review below
See attached file.

\emph{Is the treatment of the subject complete?}: No

\emph{If not, What important details / ideas/ analyses are missing?}: Omits indirect sources of overhead encountered by interpreters.
See attached file.

\emph{What type of paper is this?}: This paper is an original, previously unpublished, paper (to the best of my knowledge)

\paragraph{Recommendations to the authors}

\emph{Please list some concrete actionable items to improve the paper. When the paper is resubmitted, the authors will at least have to explain how they dealt with these recommendations.}:

I hope that the author will run the experiment comparing the performance of his interpreter to the equivalent program written in C++ and compiled with optimization. I also hope that he performs a similar experiment comparing his work to state-of-the-art Java virtual machines, for instance, and that he reports the lines of code required for his system against that of a concisely-written JIT, and also compares the lines of code required for a meta-circular JITTing VM such as Jikes/RVM

Also, work to make the writing more precise: no "This" without a noun after the word, and avoid the construct "We do ..." when you mean that your program does ....

\emph{Please help ACM create a more efficient time-to-publication process: Using your best judgment, what amount of copy editing do you think this paper needs?}: Moderate

\emph{Most ACM journal papers are researcher-oriented. Is this paper of potential interest to developers and engineers?}: Maybe

\includepdf[pages=-,pagecommand={},width=\textwidth]{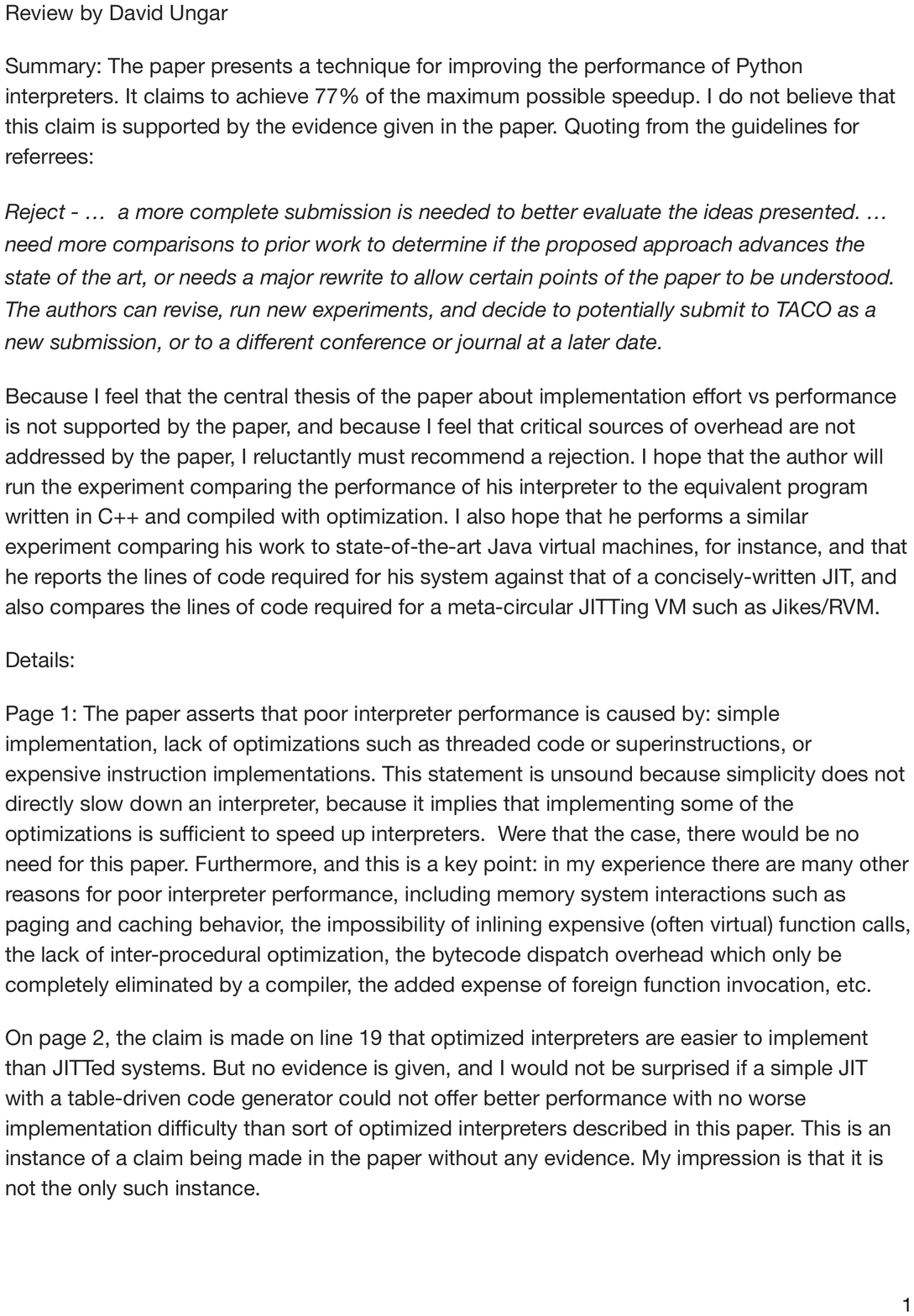}

\subsubsection{Referee 4}
\emph{Recommendation}: Needs Minor Revision

\paragraph{Comments}:
This paper proposes and evaluates an optimization ("multi-level
quickening") that works by rewriting the byte code of an interpreter,
similar to the "quickening" optimization in the JVM.  While the
original "quickening" rewrites each byte code at most once,
multi-level quickening rewrites it in two stages; also, the
optimizations are speculative, and may have to be undone (unlike
quickening).  The first optimization level, inline caching,
type-specializes individual byte codes.  The second optimization level
rewrites whole sequences of instructions; it propagates the type
information to the other instructions in the sequence using abstract
interpretation; as a result, some of the byte codes can work directly
on unboxed data, instead of unboxing and boxing data in every byte
code.

The author implemented this optimization in a Python interpreter and
presents an empirical evaluation.  One interesting feature of the
evaluation is that it takes into account and corrects for a common
problem when benchmarking interpreters: many benchmarks spend a lot of
time in library code written in another language, for which
interpreter optimizations don't help at all.  In the paper the actual
speedup is compared to the speedup that would be achieved by
optimizing the interpreted part away completely; the reported speedups
are impressive, and remarkably close to the limit in many cases.

\emph{Additional Questions}:
\emph{Review's recommendation for paper type:} Full length technical paper

\emph{Should this paper be considered for a best paper award?:} No

\emph{Does this paper present innovative ideas or material?:} Yes

\emph{In what ways does this paper advance the field?:} It presents an optimization for interpreters for dynamically typed interpreters, evaluates it, and it proves effective.  That's surprising, because earlier stuff I saw in that area were not, and I had almost given up hope.

\emph{Is the information in the paper sound, factual, and accurate?:} Yes

\emph{If not, please explain why.:}

\emph{Rate the paper on its contribution to the body of knowledge in architecture and code optimization (none=1, very important=5):} 4

\emph{What are the major contributions of the paper?:} It presents an optimization for interpreters for dynamically typed interpreters, evaluates it, and it proves effective.

\emph{It also provides an innovation in the evaluation of interpreter performance:} many benchmarks spend a lot of
time in library code written in another language, for which
interpreter optimizations don't help at all.  In the paper the actual
speedup is compared to the speedup that would be achieved by
optimizing the interpreted part away completely

\emph{Rate how well the ideas are presented (very difficult to understand=1 very easy to understand =5):} 2

\emph{Rate the overall quality of the writing (very poor=1, excellent=5):} 4

\emph{Does this paper cite and use appropriate references?:} Yes

\emph{If not, what important references are missing?:}

\emph{Should anything be deleted from or condensed in the paper?:} No

\emph{If so, please explain.:}

\emph{Is the treatment of the subject complete?:} No

\emph{If not, What important details / ideas/ analyses are missing?:} See "recommendations to the authors"

\emph{What type of paper is this?:} This paper is an original, previously unpublished, paper (to the best of my knowledge)

\paragraph{Recommendations to the authors}
\emph{Please list some concrete actionable items to improve the paper. When the paper is resubmitted, the authors will at least have to explain how they dealt with these recommendations.:} The main problem I have with this paper is that the presentation
leaves a number of questions open:

\begin{itemize}
\item - During reading sections 4.3-4.5, the connection between the
  high-level stuff in section 4.2 and the low-level stuff in section
  4.3 was too thin for my taste.  There were some explanations
  missing; e.g., I would be unable to implement the unboxing cache
  from the description.

\item   - What is the benefit of using two optimization levels?  Could you not
  just directly optimize to the NAMA level?

\item   - One problem I have noticed when considering optimizing languages
  like Python and Ruby is that Integers can turn into BigInts on every
  arithmetic operation, invalidating any followwing type
  specialization.  I do not see this problem addressed in the paper
  (indeed the addition example on page 4 ignores it).

\item   - Section 4.7 is very short.  Maybe you should give an example.
\end{itemize}

\paragraph{Detailed comments}

\begin{itemize}
\item In Figure 5, for E50 MLQ+UC+SI provides a speedup that's higher than
  the limit.  Please explain this result.

\item   I am wondering whether you should present your speedup data in another
  way in addition: Speedup of the interpreted part only (of course, this
  would produce funny results for E39 and E50), or alternatively, the
  remaining portion of time in the interpreter compared to the original
  time (which would be the inverse of the speedup).

\item   The Leone and Lee paper did not use run-time feedback as far as I
  remember.  Instead, it simply compiled code with currying such that
  passing a parameter would run-time code generate a version specialized
  for that parameter.

\item   Proof-read the references carefully; I have not checked everything,
  but the Wang, Wu, Padua 2014 reference looks incomplete.
\end{itemize}

\emph{Please help ACM create a more efficient time-to-publication process: Using your best judgment, what amount of copy editing do you think this paper needs?:} Heavy

\emph{Most ACM journal papers are researcher-oriented. Is this paper of potential interest to developers
and engineers?:} Yes

\newpage
\section{Reviewer Feedback}\label{s:reviewer-feedback}
Most of the information presented in the reviews is provided as is and the reader is free to interpret them to their liking.
For the archival purposes intended for this version of the paper, I merely want to address three specific issues identified by the reviewers:
\begin{enumerate}
\item Incorrect paper citation.
\item Requested focus on indirect branch prediction analysis.
\item Missing data on implementation efficiency.
\end{enumerate}

\subsection{Incorrect Paper Citation}
Reviewer 2 from the TACO'14 submission (see Section 3.3.4) commented on the paper falsely citing the Williams et al. paper from CGO 2010.
I remember that some details were missing in the paper, which led me to look for the corresponding PhD thesis to complement the missing information.
My entry in the corresponding Bibtex file confirms my memory:
\begin{quote}
20111018/sbr: this is unfortunately incomplete, because it does not include the technical details that were presented in williams' phd thesis, where he describes the use of a background thread that starts the system compiler (gcc) to generate the new instructions and link them in
\end{quote}
Since this paper the only one co-authored by Kevin Williams in my Bibtex file, I must not have inferred that what I read in the thesis is part of multiple separate papers.
Doing thorough related work is important and I do make this an important point for all of my graduate students.
Apologies for this error, even ten years ago this should not have happened.

\subsection{Indirect Branch Prediction Analysis}
Reviewer 2 from the TACO'14 submission (see Section 3.3.4) has another important remark on the importance of branch prediction.
Even in today, in August 2021, I am still stunned by this remark, primarily for two simple reasons.

First, the reviewer states that the model of prior work clearly shows importance of branch prediction for interpreters and that all of the speedups could also ``depend on the luck of how the C compiler lays out the interpreter code.''
Somehow this remark completely misses the fact that several interpreters with focus on optimizing branch prediction, e.g., by way of applying some form of threaded code, did \emph{not} experience the same amount of speedups.
In a mailing list, the Python implementers stated that applying threaded code (called ``computed gotos'' there) only brought about 14\% speedups.
Similarly, the paper by Vitale and Abdelrahman from 2004 showed that applying threaded code could also lead to slow-downs for some programs.

As a result of these observations, I fundamentally oppose the reviewer's view: ``Given that indirect branch prediction appears to be the most common reason for the effectiveness of interpreter optimizations, [...]''.
Optimizing branch prediction is effective only for low abstraction-level interpreters, such as those for Java, Forth, and OCaml.
All studies I have read that tried to replicate this success for high abstraction-level interpreters, such as JavaScript, Python, Perl, Tcl, failed to achieve reported speedups.

Second, lucky placement by the C compiler is highly unlikely to make up the reported speedups, as evidenced by almost identical speedup profiles for both PowerPC 970 and Intel Nehalem i7-920 architectures (see Figure 5 in the paper).
The Intel CPU uses a two-level branch predictor but it only results in little improvements over the PowerPC's branch predictor.
What Figure 5 does show, and is also explained in Section 4.6 on superinstructions, is that instruction dispatch does become a performance bottleneck when other, dominating overheads have been eliminated (compare, for example, the last two light-gray bars).

\subsection{Missing Data on Implementation Efficiency }
The paper did not contain any data on implementation efficiency.
Since David Ungar was kind enough to sign his review, I contacted him afterwards to address some of his comments.
I did measure all of the following artifacts and provide them here for reference.

\subsubsection{Collected Raw Data}

\begin{table}[t!]
  \begin{subtable}[t]{.3\linewidth}
    \centering
    \begin{tabular}[h!]{l r}
      \toprule
      Component & LoC (Java) \\
      \midrule
        Baseline\footnote{(for IA32 + PPC support, each about 5k)} & 16,600 \\
        Common &    4,357 \\
        Opt &      67,785 \\
      \bottomrule
    \end{tabular}
    \caption{Jikes version 3.1.3, in package \texttt{org.jikesrvm.compilers}.}
    \label{t:loc-jikes}
  \end{subtable}%
  \qquad
  \begin{subtable}[t]{.3\linewidth}
    \centering
    \begin{tabular}[h!]{l r}
      \toprule
      Component & LoC (RPython) \\
      \midrule
        Metainterp &  47,225 \\
        Backend &     32,125 \\
        Codewrite &    8,223 \\
      \bottomrule
    \end{tabular}
    \caption{PyPy version 3 beta, in directory \texttt{pypy/rpython/jit}.}
    \label{t:loc-pypy}
  \end{subtable}%
  \qquad
  \begin{subtable}[t]{.3\linewidth}
    \centering
    \begin{tabular}[h!]{l r}
      \toprule
      Component & LoC (C++) \\
      \midrule
     JIT compiler backends\footnote{For arm, arm64, mips, x64, ia32 architectures.} & 30,000+ \\
     Template JIT & ~38,345 \\
     Hydrogen Opt. JIT & 25,416 \\
      \bottomrule
    \end{tabular}
    \caption{Google's V8/Crankshaft JIT compiler.}
    \label{t:loc-v8}
  \end{subtable}
  \caption{Implementation effort in lines of code for Jikes, PyPy, and V8.}
  \label{t:loc-comp}
\end{table}

\Cref{t:loc-comp} lists implementation efforts in terms of lines of code required for three separate systems with comparable aspects.
\Cref{t:loc-pypy} lists data in the corresponding directory, with \texttt{pypy/pypy/interpreter}, corresponding to the interpreter, using another 27,018 lines of Python code.
\Cref{t:loc-v8} lists data for Google's V8 JavaScript engine.
Since there are many parallels between JavaScript and Python, I also downloaded and measured against. Unfortunately, the structure of the C++ code is not ideal for
measurement, so the presented data is merely an approximation.

\begin{table}[t!]
  \centering
  \begin{tabular}[h!]{l r l}
    \toprule
    Component & LoC & PL\\
    \midrule
    C-code generator & ~ 3,000 & Python \\
    C-code templates for the code generator & ~ 3,000 & C \\
    Abstract interpreter & 1,000 & C \\
    More efficient instruction format & 1,000 & C \\
    \bottomrule
  \end{tabular}
  \caption{Implementation effort in lines of code of multi-Level quickening (Python 3).}
  \label{t:loc-mlq}
\end{table}

\Cref{t:loc-mlq} lists the raw data for the multi-level quickening system described in the preceding paper.
The code generator component generates about 15,000 lines of C code for the interpreter dispatch loop and
another 12,000 lines of C code for its superinstructions.
The generated code is trivial, highly regular and thus amenable to simple code generation.



\subsubsection{Interpretation of Implementation Efficiency Data}

Jikes and PyPy/V8 are not really comparable, primarily because of differences in the programming
language they optimize. Python and JavaScript are much more like Smalltalk and Self, particularly
from an implementation perspective, as they all are more dynamic than Java.

For comparison, I am going to assume that MLQ requires 6,500 lines of code (5,000 lines of C code
plus 1,500 lines of Python code). This system is nowhere near production, and many things could be
implemented more efficiently.

\begin{table}[h!]
  \centering
  \begin{tabular}[h!]{l r}
    \toprule
  Jikes baseline: (baseline + common) / MLQ & 3.22 \\
  Jikes opt: (opt + common) / MLQ           & 11.10 \\
  PyPy: (metainterp + backend) / MLQ        & 12.21 \\
  PyPy: interpreter / MLQ                   & 4.16 \\
  V8/CS: (template JIT + backend) / MLQ     & 10.51 \\
  V8/CS: (hydrogen JIT + backend) / MLQ     & 8.53 \\
    \bottomrule
  \end{tabular}
  \caption{MLQ relative to other systems.}
  \label{t:comparison}
\end{table}

\Cref{t:comparison} contains the data comparing the control group with the MLQ system.
There are several shortcomings and reservations I have with such a comparison, but it is the easiest and quickest thing I can do right now.
In my opinion, the MLQ system compares favorably to the other systems in terms of implementation complexity.
Note that I expect that a full-blown adaptive JIT compiler will, depending on the evaluation conditions, show better peak performance than the MLQ interpreter.

\newpage
\section{Conclusions}\label{s:conclusions}
Ten years appears to be a long time; upon re-reading and remembering the parts of the story, I noticed that it was not that long after all.
There are, I believe, two sets of conclusions to be drawn from the archival version of this paper: (i) an objective assessment of the contributions, and (ii) my subjective takeaways from the experience.

\subsection{Contributions}
The paper makes the following contributions, in no particular order:

\paragraph{General mechanism} MLQ provides a general mechanism for incremental, iterative refinement of interpreter instructions based on continuous rewriting.
  Quickening in the original sense of the JVM (see Lindholm and Yellin, 1996) replaces a single instruction with another single instruction.
  Continuous replacing on the same semantic level, for example, to collect profiling information or branch frequencies, differs from the presented multi-level quickening in the following ways.
  First, such information is mostly on the single-instruction level, so no larger set of instructions will be rewritten.
  Second, the observed information is not used to optimize with increasingly specialized instructions.

  A different perspective is that profiling information or branch frequencies collect \emph{meta} information, whereas a native machine-type specialized instruction collects information specifically geared towards a specific instruction occurrence.
  Note that the MLQ implementation also uses a quickening-based approach to collect loop frequencies to determine whether or not code should be optimized.

  The MLQ approach gives way to multiple successive refinements, where implementers can choose which overheads to address to further improve performance.
  An idea, for example, that I started implementing in 2014, but never gotten around to complete, is to use MLQ to develop a hybrid instruction-set architecture interpreter that combines a stack-based interpreter instruction set with a register-based instruction set.
  The stack-based instruction set would be the default instruction set, providing space efficiency whilst incurring higher instruction dispatch costs.
  Once a function qualifies for optimizations through profiling, one could use a register-based interpreter instruction set that trades of increased space requirements for reduced dispatch costs.

\paragraph{Type-based superinstructions.}
  Prior work on superinstructions focuses on superinstructions of a pre-determined length, say superinstructions comprising four single instructions.
  In a high-abstraction level interpreter, length-based superinstructions are mostly useless, as the performance penalty is not due to instruction dispatch, but due to costly interpreter instruction implementations.
  Once MLQ successfully addressed these overheads---costly interpreter instruction implementations---known instruction dispatch optimization techniques, such as threaded code and superinstructions can unfold their full potential.
  When MLQ detects frequent patterns, however, their length is based on native-machine data types that are ill suited for deconstruction into frequently occurring pattern ``tiles.''

  An important advantage of using type-based superinstructions is that the full optimization potential of the ahead-of-time compiler used to compile the interpreter (called the \emph{staging step} in the paper) is applied.
  Since the complexity of the interpreter-instruction implementation is now at the native-machine level, incorporating information about native-machine data representations, MLQ exposes much more ``surface'' for compiler optimizations than when used with high-level interpreter instructions.
  In other words, just sticking together $n$ Python interpreter instructions and compiling them with a C compiler will not result in large performance benefits, whereas the compilation of a superinstruction comprising $n$ type-specific low-level instructions can produce native-machine code that even an optimizing JIT compiler may not apply due to latency.

\paragraph{Evaluation methodology.}
  The paper investigates the sources of interpretative overheads to examine how much optimization potential an interpreter can unfold in theory.
  Prior work, and for that matter, also most other work I am aware of as of time of this writing (August 2021), just use a set of benchmarks, run and time them, and report the results.
  With the MLQ paper, I was interested in finding a conclusive answer of the varying optimization potential experienced in well known benchmarks.
  Only for benchmarks dominated by the interpreter, optimizing the interpreter will yield benefits.

  Note that the same observation pretty much holds for most other benchmark suites.
  In SPEC, for example, I most often find that \texttt{xalanc} is most indicative of real-world behavior of performance optimizations and/or applied security techniques.
  More investigation into what \emph{actually} dominates performance is needed and certainly warranted to really understand what is going on.
  Such an investigation is usually not glamorous, and most scientists would probably not care, or maybe not even see a contribution in it, but it is important nevertheless.

\paragraph{Purely-interpretative focus.}
  In MLQ, all compilation is done at interpreter compile-time, meaning \textbf{no} code will ever be dynamically emitted.
  This purely-interpretative way of optimizing interpreters has the following advantages.

  First, one can piggyback onto an existing ahead-of-time compiler's backend, meaning one can have a reasonably fast baseline performance without requiring the complex and error prone task of creating a new compiler backend.
  Second, one need not have executable and writable memory privileges, thereby decreasing attack surface and increasing interpreter security.
  (Note that, recently, Microsoft has experimented with turning off the V8 JIT compiler in the MS Edge browser for the same reason.
  For some more data, I refer the interested reader to the corresponding blog \url{https://microsoftedge.github.io/edgevr/posts/Super-Duper-Secure-Mode/#project-super-duper-secure-mode})
  Third, an optimizing interpreter, exemplified by the MLQ system, could also be directly put into hardware.
  If, for example, on were to ``burn'' the code onto a ROM, then one could decrease energy consumption whilst increasing security.
  This scheme could also be using in slightly altered scenarios, namely in FPGAs that could adapt their optimization-level according to shifting needs, for example in a data center.
  Another exemplary scenario would be to provide computation on memory chips, where one could distribute Python code to memory chips, instead of transferring and manipulating data on the CPU.
  Fourth, the general MLQ infrastructure allows for cross native-code modules optimizations, i.e., a library, such as \texttt{numpy} could provide its own optimized instructions and/or instruction sequences, and the hook them into the optimizing interpreter.
  This aspect combines the general mechanism with the purely-interpretative nature of MLQ, and offers an interesting alternative to often problematic use of foreign-function interfaces in JIT compilers, which often limit optimizations to code that has been emitted by JIT compilers themselves.
  (In fact, I have started working on a prototype implementation of that system in 2012, but due to lack of academic incentives focused my attention on my work in language-based security.)
  Fifth, the described MLQ system poses new and exciting opportunities for interpreter code generation.
  The present system uses an unsophisticated template-based code generator producing C code, written in Python.
  An improved system could leverage the insights from the Tiger system, combined with the rewriting and type-refinement depending on the type system of a programming language and ``derive'' the proper optimized interpreter instructions.
  Sixth, bypassing dynamic code-generation enables us to formalize and verify the semantics preserving nature of the optimizations.
  (Which I did with a grad student and published in early 2021, but have tried and failed to publish at POPL'14, POPL'15, and ECOOP'17.)

\subsection{Personal Conclusions}

My personal conclusions of both conducting the research and failing at publishing its results are as follows.
The former part, conducting the research, was and still is fun.
I strongly believe that we have not yet reached the maximum possible performance of what can be achieved in single-threaded program performance, and very much so in interpreter performance.
(Though my gut feeling is that there won't be another 4$\times$ improvement to be had, more like a 2$\times$---that's at least what I expect my next optimization idea will yield.)

My failure of publishing the MLQ-system, including consideration of compounding factors, resulted in my effectively leaving the SIGPLAN community.
My aim at publishing this paper was pure vanity: Having a single-authored paper at PLDI would increase my chances for faculty applications, at least that's what I had hoped for.
These hopes were not only dashed by the negative reviews, but much more by the 2013 PLDI PC chair (Cormac Flanagan), who stated in his opening address that the least likely chance for acceptance to PLDI was research in compiler optimization.
These facts indicated that interest interpreter optimization had become out of fashion.
If I had read the signals right, or if some of the reviewers had pointed this out to me, then I would have submitted this paper to CGO instead, but nobody did and my vanity blinded me.
At the same time, our research in language-based security, particularly in software diversity took off.
The security reviewers liked our work and I felt that what we did there, was equally important to my work in interpreter optimization.

Scientific research is driven by fashion to a surprising degree.
When important topics become out of fashion, some members of the scientific community move on, while others stay and continue research.
Looking back over the past ten years, scientists who move on to greener pastures, often picking low-hanging fruit, seem to do better on bibliometric scales, i.e. publishing more and acquiring more citations.
Scientists who continue to stay in their original research areas then often find themselves between chairs: conferences change their focus to accommodate new research areas, thus altering the corresponding communities.
I did not understand this social component of CS publishing ten years ago and it still seems strange to me.
Ten years ago, a change of communities was most welcome, although this required a change of focus, I could still remain true to my interests in programming language design and implementation.
Now, in 2021, the signals in the systems security community closely resemble the situation I found myself in ten years ago---maybe now would be a good time to change back to programming languages again?

\end{document}